\documentstyle[preprint,aps]{revtex}

\def\beq{\begin{equation}}
\def\eeq{\end{equation}}
\def\beqa{\begin{eqnarray}}
\def\eeqa{\end{eqnarray}}
\def\nn{\nonumber}

\begin{document}

\title
{QUANTUM CORRECTIONS AND EXTREMAL BLACK HOLES}
\author{G. Alejandro $^1$, F. D. Mazzitelli $^{*,2}$ and
C. N\'u\~nez $^1$}
\bigskip
\address
{$^1$~~Instituto de Astronom\'{\i}a y F\'{\i}sica del Espacio\\
C.C.67 - Suc. 28 - 1428 Buenos Aires, Argentina\\
Consejo Nacional de Investigaciones Cient\'{\i}ficas y T\'ecnicas \\
  and Universidad de Buenos Aires}
\address
{$^2$~~International Centre for Theoretical Physics\\
P.O. Box 586 - 34100 Trieste - Italia}

\date{\today}
\maketitle
\begin{abstract}
We consider  static  solutions  of two dimensional dilaton gravity models as
toy laboratories to address the question of the final fate of black holes.
A nonperturbative correction to the CGHS potential term is shown to lead
classically to an extremal black hole geometry, thus providing a
plausible solution to Hawking evaporation paradox.
However,
the full quantum theory does not admit an extremal solution.
\end{abstract}
\vskip3cm
{\small * Permanent address: Departamento de F\'\i sica,
FCEyN, Ciudad Universitaria, Pabell\' on I, 1428 Buenos Aires -
Argentina and IAFE, C.C.67- Suc. 28 - 1428 Buenos Aires - Argentina}
\newpage
{\it 1.} The sound conjecture
that quantum effects are able to cure
the diseases of classical general relativity, such as singularities and the
black hole evaporation paradox, has been recently disproven in certain
cosmological \cite{mazzir,mazzig}, black hole \cite{dA,Strom} and
colliding-wave \cite{klim}
solutions to two dimensional dilaton gravity models coupled to
conformal matter fields. On the contrary,
according to \cite {mazzir,mazzig},
when the classical matter content is below a given threshold,
rather the opposite
may happen. Namely, certain nonsingular
classical cosmological  solutions  develop quantum singularities
in the weak coupling region, which suggests that they will not be removed in
the full quantum theory.  Moreover,  the  threshold and the leading behavior
of the scale factor near the singularity are independent of the couplings.
Similarly, classical solutions
describing gravitational scattering of
matter wave-packets develop curvature singularities
vanishing only in the limit
$\hbar \rightarrow 0$ \cite{klim}.

In  the  black  hole context,
the original model proposed by
Callan, Giddings, Harvey and Strominger (CGHS) \cite{cghs}
discusses Hawking radiation semiclassically  by adding the
conformal anomaly term, but cannot be solved exactly.
The full quantization of the model was addressed by a number
of authors \cite{dA,Strom,BC}. The method followed restricts the effective
action to a $c=26$ conformal field theory that reduces to the semiclassical
CGHS model in the weak coupling limit.
It turns out to be possible to solve exactly the equations of motion
of this conformal invariant theory. However,
the model leads to  unending thermal radiation and, consequently negative
Bondi mass.
This problem is related to the
fact that the rate of Hawking radiation in these models does not change
as the energy of the black hole is depleted. So, again, quantum
effects seem not useful.

Two different
proposals have been advanced in order to
overcome this problem.
The first one, originally
suggested by Russo, Susskind and
Thorlacius (RST) \cite{rst}
consists of a modification
of the CGHS effective action which admits analytic solutions.
The modified model is again a conformal field theory, but
now one of the fields has a restricted range of values
$X\geq X_{cr}$. RST proposed a set of
boundary conditions at   $X = X_{cr}$ which ensure
a finite spacetime curvature and also stop Hawking radiation.
The RST boundary conditions have been subsequently
criticized \cite{stromtor}, because
they are incompatible with natural quantum mechanical
boundary conditions for the matter fields. As an alternative,
it has
been proposed that boundary conditions   should be imposed
on all fields along a timelike curve,
the analog of the origin of radial coordinates in four
dimensional gravity \cite{stromtor}.

Here we will be concerned with
a completely different solution to the unending Hawking
radiation. As proposed by Banks and O'Loughlin (BO) \cite{banks},
if the two dimensional theory is modified in order to admit
black holes of
Reissner Nordstrom type,  then, in the extremal limit,
the temperature vanishes and a stable end point
is achieved.
The extremal black holes proposed by
BO \cite{banks}
are plausible remnants but, unfortunately,
the models are difficult to handle and
they have not been analytically solved at the semiclassical level
(see also \cite{lo}).

Following BO's idea, we will modify the CGHS action by adding
a nonperturbative
correction to the potential term which allows classically an extremal
black hole solution.
We will show that it is possible to construct
the conformal field theory associated to this model
and thus perform the quantization procedure \`alla David,
Distler and Kawai \cite{ddk}. Even though this conformal field
theory is not
easily analytically solvable,
it is possible to show that the extremal geometry
is not a solution of the full quantum system.
Thus, we will conclude that this quantization procedure
spoils the hope to
stop Hawking radiation by this means.

In Section 2 we perform the quantization of an arbitrary
dilaton gravity lagrangian
by generalizing the method introduced by
 de Alwis (dA) \cite{dA}.
In Section 3,
the unending evaporation problem of two dimensional black holes is
discussed and one particular solution is analyzed both
classically and quantum mechanically. Conclusions are presented
in Section 4.

\bigskip


{\it 2.} The most general renormalizable dilaton gravity lagrangian can be
written as
\beq
{\cal L} = \sqrt{-g} \left [
D(\phi) R + G(\phi) (\bigtriangledown \phi)^2 + V(\phi) \right ]
\label{large}
\eeq
where $D(\phi)$, $G(\phi)$ and $V(\phi)$
are arbitrary functions which,
in order to recover the
CGHS theory in the weak coupling limit, should satisfy
\beq
D(\phi)\rightarrow {G(\phi)\over 4} \rightarrow e^{-2\phi}~~, \qquad
V(\phi) \rightarrow 4 \lambda^2 e^{-2\phi}
\eeq
as $e^{2\phi} << 1$.

As shown by BO, for computational reasons,
it is convenient to eliminate the G term. This can be done in
a nonsingular way
when $D^{^{\prime}}={{dD}\over {d\phi}}\neq 0$,
by performing a conformal
transformation $g_{\mu\nu}=e^{2S(\phi)}\hat g_{\mu\nu}$
 with $2S^{\prime}(\phi) D^{\prime}(\phi) = -G(\phi)$.
Thus the lagrangian reduces to
\beq
{\cal L }= \sqrt{-\hat g}\left [ D(\phi) \hat R + W(\phi) \right ]
\label{redu}
\eeq
where $W=V e^{2S}$.
In order to quantize the system we generalize to this case
 the procedure originally introduced
 by dA \cite {dA}.
In conformal gauge $\hat g=e^{2\hat\rho}\tilde g$, the action (\ref{redu}),
supplemented with the Liouville action, takes the semiclassical form
\beq
S[\tilde g] = {1 \over {4\pi}} \int d^2\sigma
\sqrt{-\tilde g} \left \{[D(\phi)-\kappa\hat\rho] \tilde R + 2 D^{\prime}
(\phi) \tilde \bigtriangledown \hat\rho \tilde \bigtriangledown\phi -
\kappa (\tilde \bigtriangledown \hat \rho)^2 + W(\phi) e^{2\hat
\rho} \right \}
\eeq
which can be written as a non linear $\sigma$-model
\beq
I[X,\tilde g] = -
{1 \over {4\pi}}\int d^2\sigma [{1 \over 2} \tilde g^{ab}
 G_{\mu\nu} \partial_a X^\mu \partial_b X^\nu + \tilde R \Phi(X) + T(X)]
\label{sigmamo}
\eeq
where $X \equiv (\phi,\hat \rho)$, $G_{\mu\nu}$ is a metric in
the space of fields $(\phi,\hat \rho)$ and
$\Phi$ and T are the dilaton and tachyon fields respectively.
In order to define a conformal field theory,
the couplings $G, \Phi$ and T must solve
the $\beta$-function equations with the weak coupling boundary conditions
\beqa
G_{\phi\phi}=0 \qquad G_{\phi\hat\rho}=-2D^{\prime}(\phi) \qquad
G_{\hat\rho\hat\rho}=2\kappa \nn \\
\Phi = -D(\phi) + \kappa \hat\rho \qquad
T = -W(\phi) e^{2\hat\rho}
\label{bc}
\eeqa
Recall that the effective coupling is $|D^{\prime}(\phi)|^{-1}$
and thus, the weak coupling region is characterized by $|D^{\prime}|^{-1} <<
1$.

Parametrizing the target space metric as
\beq
ds^2 = h(\phi) d\phi^2 - 4 D^{\prime} [1+\bar h(\phi)] d\hat\rho d\phi +
2\kappa [1+\bar {\bar h}(\phi)] d \hat\rho^2
\label{tm}
\eeq
with $h(\phi) = O(D^{\prime 0})$ and $\bar h(\phi) = O(D^{\prime -1})
$, it can be easily shown that ${\bf R}(G_{\mu\nu}) = 0$
if $\bar {\bar h} = 0$. Therefore in this case it is possible
to find a Minkowskian coordinate system. Defining
\beqa
X &\equiv & {2\over\kappa}
\int d\phi \left [{{D^{\prime 2}}\over 4}
(1+\bar h)^2 - {1 \over 8} \kappa h \right ]^{1/2} \nn \\
Y &\equiv & \left [\hat\rho
- {1 \over \kappa} \int d\phi (1 + \bar h(\phi)) D^{\prime}(\phi) \right ]
\label{xy}
\eeqa
the target space interval reads $ds^2=2\kappa (-dX^2+dY^2)$ and
the action (\ref{sigmamo}) can be written as
\beq
I={1\over {4\pi}} \int d^2\sigma \left [-4\kappa\partial_+X\partial_-
X +4\kappa \partial_+Y\partial_-Y - T(X,Y) \right ]
\label{sigmamo2}
\eeq
The gravitational $\beta$-function
\beq
 {\beta_{\mu \nu} = {\bf R}_{\mu \nu} + 2 \nabla_\mu^G \partial_\nu
   \Phi -\partial_\mu T \partial_\nu T + ... = 0}
\eeq
is easily seen to be solved, neglecting terms of $O(T^2)$ and
imposing the boundary conditions (\ref{bc}), by $\Phi =\kappa Y$.
Replacing this linear dilaton solution in the dilaton $\beta$-function
\beq
 \beta_{\Phi} = -{\bf R} +
    4G^{\mu\nu} \partial_\mu\Phi \partial_\nu \Phi -
    4\nabla_G^2 \Phi
+{{N-24}\over 3}+
    G^{\mu \nu} \partial_{\mu}T \partial_{\nu} T - 2T^2 + ...=0
\eeq
$\kappa$ turns out to be
$\kappa = {{24 - N} \over 6}$,
N being the number of conformal fields.

T(X,Y) should satisfy the tachyon $\beta$-function, namely
\beq
\beta_T = -2\nabla _G^2 T
+ 4 G^{\mu\nu}\partial_\mu\Phi\partial_\nu T - 4T + \ldots = 0
\label{betatac}
\eeq
If we restrict to functions $W(\phi) = const
= 4\lambda^2$, a possible solution of (\ref{betatac}) to leading order, is
$T(X,Y) =- 4\lambda^2 e^{\beta X + \alpha Y}$
with $\kappa^{-1}(\beta^2-\alpha^2)+2\alpha -4=0$. In order
to match the weak coupling boundary conditions
we must have $\alpha =2,\,\beta=\mp2$, where
the upper (lower) sign corresponds to $D'<0 (D'>0)$.

We have quantized the 2D models using the reduced lagrangian
(\ref{redu}) instead of the original lagrangian (\ref{large}).
In performing the quantization  with (\ref{large}),
the target space metric would be parametrized  as
\beq
ds^2=-2G(\phi)(1+l)d\phi^2-4D'(\phi)(1+\bar l)d\phi d\rho
+2\kappa d\rho^2
\label{tmlarge}
\eeq
A comparison of Eqns. (\ref{tm}) (with $\bar{\bar h}=0$)
and (\ref{tmlarge}),
shows that
both quantization procedures are equivalent if
\beqa
h&=&-2G(\phi)(1+ l)-4D'(\phi)(1+\bar l)S'+2\kappa [S'(\phi)]^2
\nn\\
\bar h &=& \bar l-\kappa{S'(\phi)\over D'(\phi)}
\label{equiv}
\eeqa
We will use these relations in the next Section.




\bigskip

{\it 3.} Let us  now  consider the CGHS model with
an additional tachyon contribution.
 Namely,  let us take in (\ref{sigmamo2}) for $\kappa > 0$
\beq
T(X,Y)  =  -4\lambda^2    e^{2(Y-X)}    +    \mu
e^{2 (X+Y)}\equiv -e^{2\hat\rho}  W(\phi)
\label{newtac}
\eeq
which is also a solution  of  the  tachyon $\beta$-function
(\ref{betatac}) satisfying
the  weak  coupling  boundary  conditions    (\ref{bc}).        Indeed,   as
$\phi\rightarrow -\infty$, it is easily seen that
\beq
T(X,Y) =-e^{2\hat\rho}\left [ 4\lambda^2 -  \mu e^{-{4\over \kappa}
e^{-2\phi}} \right ] \rightarrow
-4\lambda^2 e^{2\hat\rho}
\label{wbar}
\eeq
(note that this analysis does not hold for $\kappa <0$).
The additive term may be viewed as  a
nonperturbative quantum correction to
$W(\phi)$  which  vanishes  very rapidly
in the weak  coupling  region,  but
introduces interesting consequences
for the black hole solutions.  Indeed, as shown by BO, models in which
$W(\phi)$
has one zero have extremal black hole solutions, and are thus plausible
remnants to solve Hawking's evaporation paradox.
Similar non-perturbative corrections have been
proposed in Ref.\cite{BB}.

Let us first consider the
reduced lagrangian
\beq
{\cal L }= \sqrt{-\hat g} \left [ D(\phi)R +  W(\phi) \right ]
\label{om}
\eeq
where $ W(\phi)=4\lambda^2-\mu e^{-{4\over \kappa}e^{-2\phi}}$
and $D(\phi)=e^{-2\phi}$.
In terms of the X, Y fields defined by Eq.(\ref {xy}) with $h=\bar h=0$,
\beq
X=-{1\over\kappa}e^{-2\phi}~~,\qquad
Y=\rho-\phi- {1\over\kappa}e^{-2\phi}~~,
\label{xyclas}
\eeq
this  lagrangian reads
\beq
{\cal L}= - 8 \kappa \partial_+X \partial_-X +8\kappa \partial_+Y \partial_-X
+ 4 \lambda^2 e^{2(Y-X)} - \mu  e^{2(X+Y)}
\label{rxy}
\eeq
The equations of motion are  given by
\beq
\partial_{+-}(Y-3X)=-{2\lambda^2\over\kappa}
e^{2(Y-X)}~~,\qquad
\partial_{+-}(Y-X)= - { \mu\over 2\kappa}
e^{2(Y+X)}
\label{cxy}
\eeq
We would like
to see if these equations admit an extremal black hole solution.
In the
neighborhood of an extremal horizon,
$ds^2=-(\alpha r- 1)^2 d\tau^2+{dr^2\over (\alpha r - 1 )^2}
=e^{2\rho(\sigma)}[-d\tau^2+d\sigma^2]$, where
$\rho(\sigma)\rightarrow -ln\alpha|\sigma|$ as $\sigma\rightarrow
-\infty$.
Assuming that near the horizon $\phi =\phi_0+g(\sigma),$ with $g(\sigma)
\rightarrow 0$, the above equations lead to
 \beq
 {1\over    \sigma^2}    =  - {8\lambda^2\over\alpha^2\sigma^2}
X(\phi_0)  + O({1\over\sigma^3})~~,\qquad
{1\over    \sigma^2}    =  - {2\mu\over\alpha^2\sigma^2}X(\phi_0)
 e^{4X(\phi_0)}+ O({1\over\sigma^3} )
\label{apro}
\eeq
It is easy to see that, to lowest order in $\sigma^{-2}$,
 $\alpha^2=-8\lambda^2 X(\phi_0),~~
 X(\phi_0) =-{1\over 4} ln ({ \mu\over 4\lambda^2})$ is a solution
(notice that, as $X<0$, the solution exists
only if ${ \mu\over 4\lambda^2}>1$).
Therefore, the equations admit an extremal horizon.

Since this was an approximate  analysis, we would like to
confirm the validity of the result.
Eqns (\ref {cxy}) imply that ${dX\over d\sigma} = C e^{2(Y-X)}$,
where $C$ is an integration constant.
As a consequence
\beq
e^{2(Y-X)}=-{1\over\kappa C^2}\int dX~~W(X)
\equiv -{1\over\kappa C^2}P(X)
\label{pp}
\eeq

A horizon is a point where $e^{2\rho}=-{1\over\kappa X}e^{2(Y-X)}
=0$, $\rho$ being the
Liouville field of the CGHS metric
related to the new metric $\hat g_{\mu\nu}$ by $\rho=\hat\rho + \phi$.
Therefore, the zeros of $P(X)$ define the horizons.
When $P$ has a linear zero the behaviour of the solutions is
exactly as in the CGHS model. When $W$ has one zero,
the function $P$ will have two. If there exists a particular value
of the ADM mass for which both zeros coincide, it would
define an extremal  black hole
\cite{banks}. Quantum matter fields in this
background would not Hawking radiate.

The function $P(X)$
defined in Eqn.(\ref{pp}) is given by
\beq
P(X)= 4\lambda^2 X  - {\mu\over 4}
e^{4X} + {A\over \kappa}
\label{p}
\eeq
where $A$ is an integration constant, related to the ADM mass of
the black hole by $M={A\over 4\lambda}$.

$P(X)$ has a maximum at $X(\phi_0)$.
A simple analysis shows that it has one simple zero when
$A > {{\mu\kappa}\over 4}$,
two simple zeros
for ${{\mu\kappa}\over 4}>A>A_{crit}=4\lambda^2 \kappa({1\over 4}-X(\phi_0))$,
a double zero for $A=A_{crit}$ and no zeros for
$A<A_{crit}$.
Therefore, the   model  we  have discussed possesses
a classical extremal
solution with mass $M_{crit}={A_{crit}\over 4\lambda}$. The
value of $X$ at the horizon is $X(\phi_0)$, as anticipated
from the approximate calculation.

It  is  well
known that  the  Reissner-Nordstrom  extremal geometry leads to null Hawking
radiation.  Applying  the  procedure  introduced  by Christensen and Fulling
\cite {chris} to compute the Hawking temperature
from the trace of the energy momentum tensor, we find that
for this model, $T$ is proportional to $X{d\over dX}\left (
{P(X)\over X}\right)\vert_{X_h}$,
where $X_h$ is the value of $X$ at the horizon nearest to the
weak coupling region. Therefore there are two qualitatively
different possibilities for the Hawking temperature, depending
on the multiplicity of the roots of $P$.
If $P$ has two simple roots, i.e.
two non coincident horizons, then
the temperature will have a non zero finite value.
If, on the contrary, both horizons coincide (extremal geometry),
then $P$ will have a double zero
and the temperature will vanish.

It is interesting to note that this behaviour holds for any model
in which $W(\phi)$ has one zero. Unfortunately, 
as we  shall  now show,  the extremal
geometry is not a solution of
the full quantum system.

In order to  do  this  we  must solve the equations of motion following from
(\ref{sigmamo2}), using (\ref{newtac}). Namely,
\beq
\partial_{+-}(X+Y)={2\lambda^2\over\kappa}
e^{2(Y-X)}~~,\qquad
\partial_{+-}(X- Y)={ \mu\over {2 \kappa}}
e^{2(Y+X)}
\label{oxy}
\eeq
where
\beq
X={2\over\kappa}\int d\phi  {\sqrt {e^{-4\phi}(1+\bar  h)^2 - {1\over
8}\kappa h}}~~,\qquad
Y= \left (\hat  \rho  -  {e^{-2\phi}\over  \kappa}  +  {2\over
\kappa} \int d\phi \bar h e^{-2\phi}\right )
\label{xxyy}
\eeq
and $\hat \rho = \rho -\phi$.

Even though the general analytical solution of (\ref{oxy}) is not easy to
find,  we will proceed with the approximate analysis performed in the classical
case,
Eqns. (\ref{cxy}).
In order to do this,
we choose Strominger's $l=-e^{-2\phi}$  and
$\bar  l= -2e^{-2\phi}$  so as  to  ensure  no  ghost
radiation \cite{Strom}.  These give, according to
Eqn.(\ref{equiv}) $h=2\kappa -  8$ and $\bar h =
e^{2\phi}({\kappa\over 2}
- 2)$.  Replacing in (\ref{xxyy}), $X$ and $Y$ turn out to be
\beq
X={2\over\kappa}\int  d\phi  {\sqrt{e^{-4\phi}  - (4-
\kappa)e^{-2\phi} +
(4-\kappa)}}~~,\qquad
Y= \rho    -    {1\over  \kappa}\left  (e^{-2\phi}  +
4\phi\right )
\label{osxy}
\eeq

As $\rho$ and $\phi$  vary  between  $+ \infty$ and $- \infty$, so does $Y$.
$X$ also shares this  property for positive $\kappa$
(note that $0< \kappa < 4$).  As a consequence,
equation (\ref{sigmamo2}) defines a conformal field  theory
without boundaries
and thus,
unless boundary conditions are imposed by hand
at a timelike curve \cite{stromtor},
the only way to stop Hawking radiation is through
 an extremal black hole
geometry.  Let us see if the equations (\ref{oxy}) admit such possibility.

Let us consider first  a  given  finite  value  of
the coupling, $\phi\rightarrow
\phi_0$, since this is where the extremal classical solution was found. Thus,
\beq
X\simeq F(\phi_0)~~, \nn \qquad
Y\simeq - ln \alpha |\sigma| + G(\phi_0)
\eeq
where $F$ and $G$ can be read from Eq.(\ref{osxy}).
Eqns. (\ref{oxy}) turn out to be  incompatible  upon replacement of these
 expressions. Indeed, they yield
 \beq
 {1\over    \sigma^2}    =   -{8\lambda^2\over\kappa}
 {e^{2\left   [  G(\phi_0)  -  F(\phi_0)\right  ]}\over
 {\alpha^2 \sigma^2}}+ O({1\over\sigma^3} )~~,\qquad
{1\over    \sigma^2}    =   {2\mu\over\kappa}
 {e^{2\left   [  G(\phi_0)  +  F(\phi_0)\right  ]}\over
 {\alpha^2 \sigma^2}}+O({1\over\sigma^3} )
\eeq
and both rhs have opposite signs, contrary to the classical case.
The difference can be traced back to the conformal anomaly.
Indeed, the difference between the Lagrangians (\ref{rxy}) and
(\ref {sigmamo2}) is the conformal anomaly term
$4\kappa\partial_+(Y-X)
\partial_-(Y-X)$. This term
 flips the relative sign
of the field $Y$  in the equations.
Therefore  the  classical extremal horizon at  finite  coupling is erased by
quantum effects. Notice that this result is general, i.e. independent
of the particular quantum correction functions $h$ and $\bar h$.
However this does not mean that the extremal geometry is
not a solution, since the double horizon may still be
allowed in the weak or strong
coupling regions, to which we now turn.

Consider  weak coupling.  As $\phi\rightarrow -\infty$,
\beq
X\simeq {1\over\kappa}\left  [-e^{-2\phi}  +  (\kappa  - 4)\phi \right
]~~,\qquad
Y  =  \rho - {1\over\kappa}e^{-2\phi} - {4\over \kappa}\phi
\eeq
Replacing in (\ref{oxy}), it is easy to see that $\rho - \phi$ is a free field
and the CGHS solution is obtained, as expected.
There is no extremal horizon at weak coupling.

The
remaining possibility is $\phi\rightarrow  +\infty$,  i.e.  strong coupling.
In this case,
\beqa
X\simeq {2\over\kappa}{\sqrt{4-\kappa}\over 2} \phi = a f(\sigma),\nn\\
Y\simeq  \left  (  \rho  -    {4\over    \kappa}\phi\right  )  =
- ln \alpha |\sigma| - f(\sigma)
\eeqa
where $f(\sigma)\rightarrow  +\infty$  when $\sigma \rightarrow -\infty$ and
$0<a=\sqrt{{4-\kappa\over4}}<1$.
Replacing in (\ref{oxy}) the equations read
\beqa
{1\over    \sigma^2}    -        (1-a)    f^{\prime\prime}    =
-{8\lambda^2\over\kappa}{e^{- 2  (1+a)  f}\over
{\alpha^2 \sigma^2}}\nn\\
{1\over    \sigma^2}    -        (1+a)    f^{\prime\prime}    =
{2\mu\over\kappa} {e^{-2  (1-a)  f}\over
{\alpha^2 \sigma^2}}
\eeqa
Thus,  in  the  limit  $\sigma\rightarrow  -\infty$
the exponentials $e^{-2(1\pm a)f}$ tend to zero
and
both  equations  are
incompatible  unless  $a=0$, i.e. the uninteresting case $N=0$,  where
there is no Hawking radiation.

Since the analysis has covered all possible coupling regions, namely $-\infty<
\phi<\infty$,
the conclusion is then that the equations  (\ref{oxy}) do not admit an
extremal black hole  solution. It is important to stress
that this result depends strongly on the particular
form of the additional tachyon contribution.
Had we considered a different potential, such as
$W(\phi)= 4\lambda^2-\mu e^{-{e^{-2\phi}\over b}}$
with $b\gg {\kappa}$, the extremal black hole
would have survived. However, $b={\kappa\over 4}$
is fixed by the requierement of conformal invariance
(vanishing of the tachyon $\beta$-function).

\bigskip

{\it 4.}
At the classical level, the theory defined by Eq.(\ref{om})
admits black holes of the Reissner Nordstrom type. One would
be tempted to argue that, in this model, the black hole
evaporation would
take place until the extremal limit
is reached.
In order to discuss this
issue we should consider the dynamical problem.
However, we have shown in the previous Section that
the extremal geometry is destroyed by quantum effects.

The example above calls for discreet  prudence  before extracting
  quick conclusions
 from classical results.  In particular,  the    belief
related to the possibility that
extremal black hole remnants solve Hawking's
evaporation paradox
might not be correct.
We  have  considered  one particular example where the non-perturbative
term in the potential was chosen to define a conformally invariant theory.
The quantum  equations of motion are qualitatively different from the
classical ones. Thus, a small modification of the classical geometry,
as considered in \cite{lo},
is not  a solution.
The extremal horizon disappears.

 Our results do not rule out extremal remnants as the end point of
black hole evaporation. They strongly
suggest that, in these models, it  is  not  always
safe to appeal to the benefit of  quantum matter corrections, before
including quantum gravitational and dilaton effects.

{\it Acknowledgments:} This research was supported by
CONICET, Universidad de Buenos Aires
and Fundaci\' on Antorchas. F.D.M. and C.A.N. would like to
thank Prof. A. Salam, IAEA and UNESCO
for hospitality at ICTP.

\end{document}